\documentclass{moriond}

\def\be{\begin{equation}}
\def\ee{\end{equation}}
\def\bea{\begin{eqnarray}}
\def\eea{\end{eqnarray}}

\newcommand{\Photo}
{\includegraphics[height=35mm]{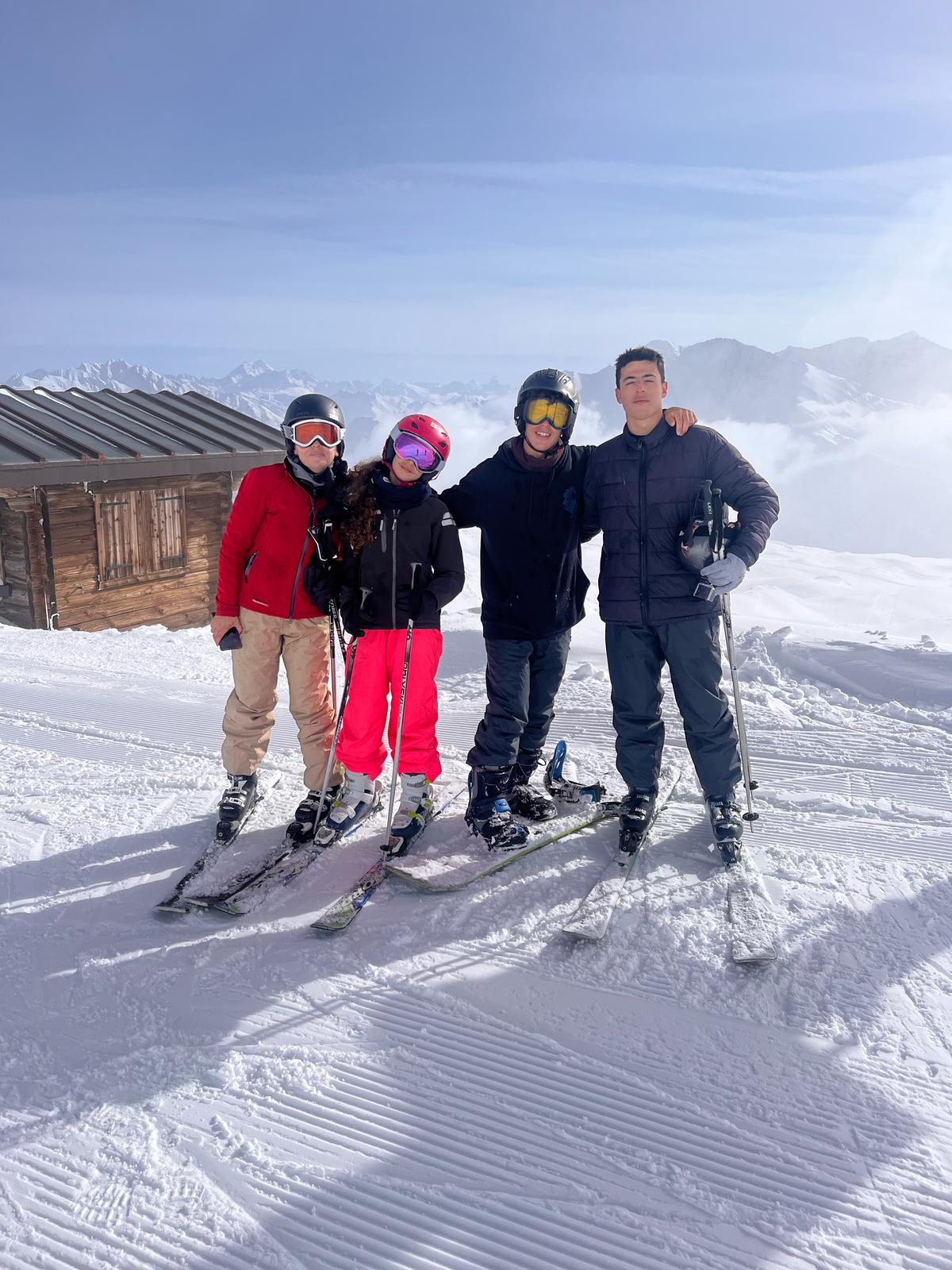}
\includegraphics[height=35mm]{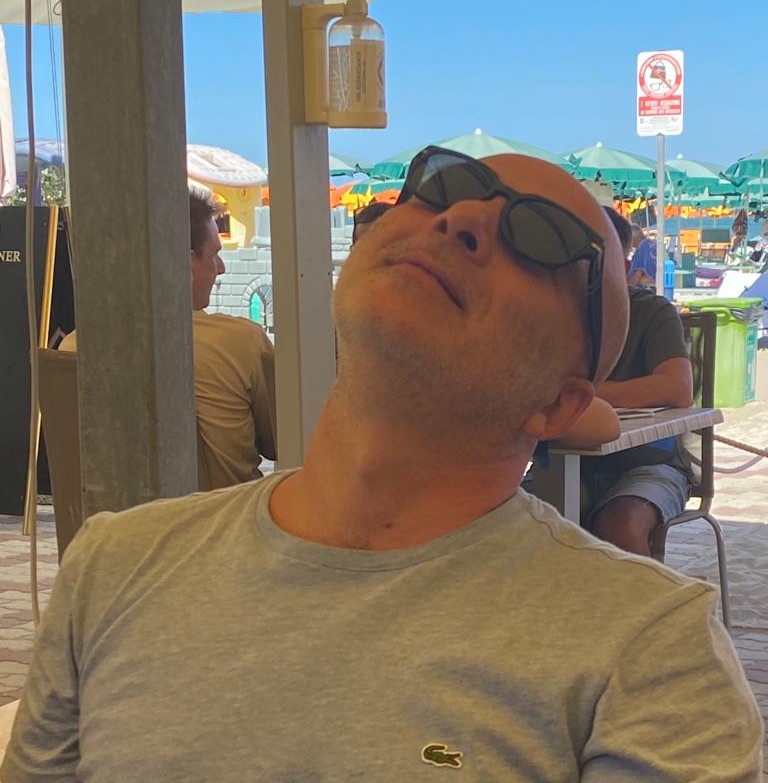}}

\begin{document}
\vspace*{4cm}
\title{Hunting for newborn magnetars: a multi-messenger approach}

\author{Dafne Guetta and Simone Dall'Osso}
\address{Physics Department,Ariel University, Ariel, Israel\\
INAF - Istituto di Radioastronomia, via Gobetti 101, Bologna, 40129, Italy}


\maketitle\abstracts{
 We carry out a numerical calculation of magnetar-powered shock break-outs (SBOs) and supernova (SN) light-curves. In particular, we investigate the impact of gravitational wave (GW) emission by the magnetar central engine on its electromagnetic (EM) counterparts in the ULTRASAT band. Our results show that GW emission by the magnetar has only a minor effect on the SBO light-curve. However, we find that SN light-curves can carry a direct signature of GW emission, which becomes more evident at late times ($> 20-30$ days).~Our results demonstrate that future ULTRASAT observations will provide crucial insights into the magnetar formation process, and unique information for direct searches of long-transient signals with current and future generation GW detectors. In particular, we estimate a rate of multi-messenger (UV+GW) detections of newly formed magnetars $>$ 1 every two years with ULTRASAT and the Einstein Telescope.}

\section{Introduction}\label{sec:introduction}

The study of astrophysical transients associated to  cataclysmic events has been a major focus of time domain astronomy in the last decade, with many dedicated surveys (e.g. ZTF, EUCLID, eROSITA, VRO).~Several such transients are expected to have a significant emission component in the UV band \cite{sagiv2014science,kulkarni2021science}.~The Ultraviolet Transient Astronomy Satellite (ULTRASAT; \cite{ultrasat}) is a small mission dedicated to time domain observations in the NUV band; scheduled to launch in 2027, it will carry out the~first wide-field survey of transient and variable sources, with a much enhanced sensitivity with respect to previous UV detectors.~One of the main science cases  
for ULTRASAT is the early~(hrs) detection of core-collapse supernovae~(CCSN) and the high cadence (minutes) monitoring of their light curves.~Observational constraints suggest that magnetars are formed in roughly 10\% of CCSNe\cite{gaensler2005,beniamini2019formation} and, based on theoretical arguments, these highly magnetic neutron stars (NS) are thought to be formed  with millisecond spin periods and huge interior magnetic fields ($>$ several $\times 10^{15}$~G).~Due to these extreme conditions, a newly formed magnetar will be able to release in the surrounding environment a large amount of energy and
are considered to be prime candidates for multi-messenger astrophysical studies of NS \cite{cutler2002gravitational,dall2021millisecond}.

In \cite{Sandhya23}  we considered magnetar-driven shock break-outs (SBOs), a specific signature of newly born magnetars particularly relevant in the ULTRASAT band.~In that work we adopted approximate power-law solutions for the evolution of the SBO physics properties (e.g., radius, velocity, pressure), and corresponding approximate expressions for the expected SBO luminosity, in order to estimate their detectability with ULTRASAT.~ In this work we extend that study to calculate exact numerical solutions for the SBO evolution and its luminosity, which also allow for the inclusion of further physical mechanisms that were previously neglected.~In particular, we account for the possibility of significant GW emission from newly born magnetars, which will reduce the available energy in the EM channel, directly impacting the shock dynamics and its observational signatures, e.g. the SBO.~Additionally, we calculate the supernova (SN) light-curve that follows the SBO and that is similarly affected by energy injection from the magnetar central engine and by GW emission.~

GW signals from newly born magnetars are expected to be stronger than those from the core-collapse itself, making them detectable at greater distances (\cite{dall2021millisecond} and references therein).~This emphasizes the importance of directed searches for magnetar-powered GW signals, particularly with the future O5 science run of advanced GW detectors (LVK), as well as the planned third-generation detectors (Einstein Telescope and Cosmic Explorer). SBOs occur prior to the SN, in the earliest stages of the explosion evolution, and will thus serve as unique triggers fors for directed GW searches (as well as for other EM follow-up observations), which can contribute to enhancing the search sensitivity. 

\section{Magnetar-driven SBO light curves and Effects of GW emission}
\label{sec:modelling}
A newly born magnetar, with angular velocity $\Omega=2\pi/P$ and moment of inertia $I$, has the rotational energy 
\begin{equation}
   E_{\rm m} = \frac{1}{2} I \Omega^2 \approx 3 \times 10^{52}~{\rm erg}~P^{-2}_{\rm ms} M_{1.4} R_{12}^2 \, ,
   \label{eq:1}
    \end{equation}
 with the spin period in milliseconds, the mass in units of 1.4 M$_{\odot}$ and the radius $R=12$ km.

The NS spins down due to magnetic dipole radiation on a timescale $t_m$ (e.g., eq. 2 in \cite{Sandhya23}), 
releasing rotational energy at the rate $L_m$.
The released energy inflates a high-pressure bubble of relativistic particles and magnetic field that sweeps the  previously launched SN ejecta into a thin shell of radius $r_s$, driving a shock through it.

In \cite{Sandhya23} we assumed that the magnetar spins down only due to magnetic dipole radiation.~In this work we take into account the effect of an additional torque, i.e.  W emission, contributing to the magnetar spindown. We thus turned to a numerical solution of the mass, momentum, and energy equations in the shock \cite{Cheva05}, in order to determine the exact time evolution of the parameters that affect the SBO luminosity, with and/or without GW emission.

The birth parameters of magnetars are still widely discussed in the literature.~The currently leading scenario suggests that
their strong $B$-field is a consequence of very fast rotation at birth - a few milliseconds at most - which sustains a powerful dynamo action \cite{1992Natur.357..472U}.~The combination of a very strong magnetic field and extreme rotation implies favorable conditions for strong GW emission, as well as for powering bright EM transients. 

\begin{figure*}[ht]
    
    \includegraphics[scale=0.3]{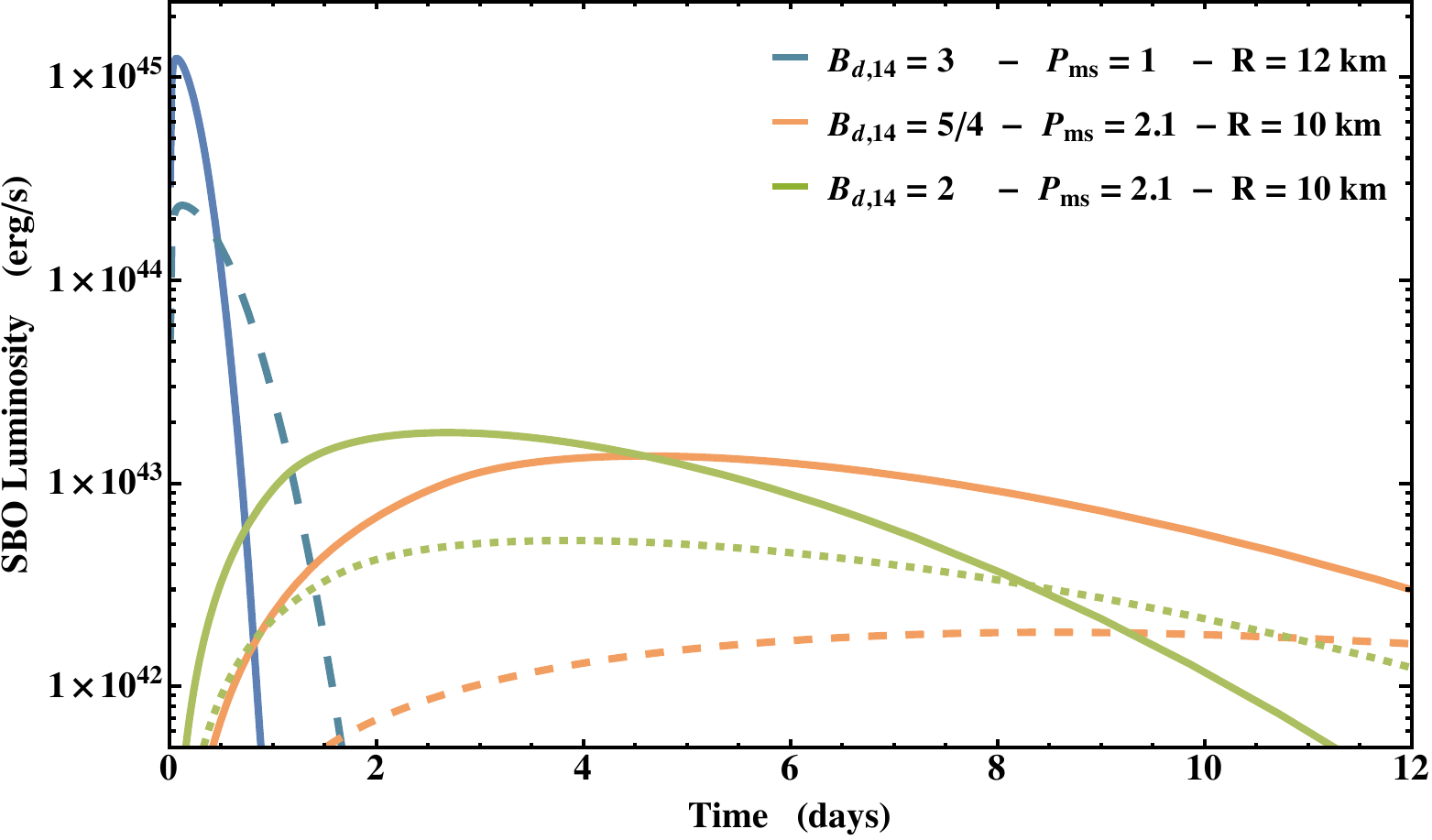}
   \includegraphics[scale=0.3]{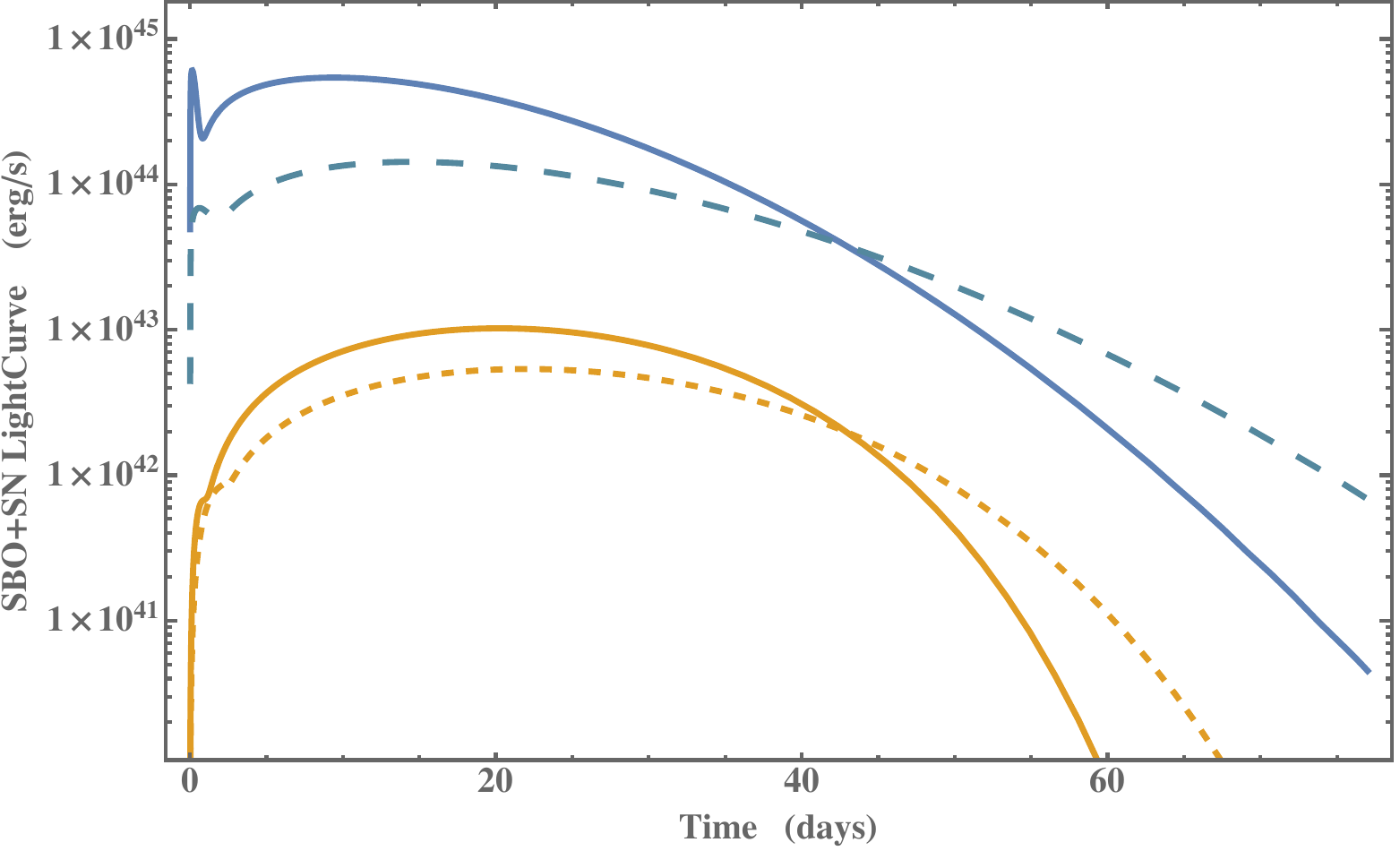} 
        \caption{{\it Left Panel:} Bolometric SBO light curve~The dashed curves next to each of the solid curves are obtained for the same magnetar parameters, adding the emission of GW with an ellipticity $\epsilon = 2\times 10^{-3}$.
        {\it Right Panel:}
        The composite light-curves for the SBO + SN, bolometric (upper, blue) and in the ULTRASAT band (lower, orange) for $z=0.1$, in the case with (dashed curves) and without (solid curves) GW emission ($\epsilon=2\times 10^{-3}$).~{\it Left Panel}: curves are drawn for $B_{d, 14} =2$ and P$_{\rm ms} =1$, and fiducial values for the other parameters }
        \label{fig:my_label1}
\end{figure*}

GW emission is caused by a deviation of the NS from spherical symmetry, which is parametrized in terms of a NS ellipticity, $\epsilon$.~GWs drain energy from the magnetar's spin, and thus reduce the available power in the EM channel, i.e. the rate of energy injection in the shock.~For a maximally-emitting NS with ellipticity $\epsilon$, moment of inertia $I$ and spin period $P$, the GW luminosity is
\be
L_{GW} = \frac{32}{5} \frac{G}{c^5} (I\epsilon)^2 \Omega^6 \, ,
\ee
and the total spindown luminosity of the NS is 
\be
\label{eq:difftot}
I \Omega \dot{\Omega} = L_{EM} + L_{GW} = -\frac{\mu^2 \Omega^4}
{c^3} - \frac{32}{5} \frac{G}{c^5} (I\epsilon)^2 \Omega^6 \, .
\ee
By solving Eq.~\ref{eq:difftot} for $\Omega$, we calculate the energy injection power, $L_{EM}$, and the corresponding evolution of the shock parameters with and/or without GW emission included.  
the left panel of Fig.~\ref{fig:my_label1} illustrates the time evolution of  SBO bolometric luminosity for fiducial values of the SN ejecta ($M_{\rm SN} = 5 M_\odot$) and explosion energy ($E_{\rm SN} = 10^{51}$ erg), for three different combinations of the NS spin and magnetic dipole field, and for the case with (dashed curves) and without (solid curves) GW emission (fixed $\epsilon=2\times 10^{-3}$).
The right panle of Fig.~\ref{fig:my_label1} shows the predicted composite light-curve in the ULTRASAT band from the SBO+SN for two different combinations of the NS spin and magnetic dipole field, $B_d$, with (dashed curves) and without (solid curves) GW emission (fixed $\epsilon=2\times 10^{-3}$).~The lightcurves were calculated for a source at $z=0.1$.~The plots show that the SBO signature  is clearly distinguishable from the SN lightcurve, and that the impact of GW emission on both SBO and SN can be significant over a wide parameter range.

\section{Results and Conclusions}

\label{sec:conclusions}

Our results illustrated in Fig.~\ref{fig:my_label1} (Right Panel) provide several insights. A striking feature is the clear separation between the SBO and SN peaks in the bolometric lightcurves, implying these two components are clearly distinguishable, over an interesting range in parameter space.~In the UV band, the SBO peak becomes much less prominent, as a result of the different evolutions of the emission spectra of the SN and the SBO.~Taken at face value, this suggests that broadband follow-up observations of ULTRASAT sources will offer the best opportunity for identifying the SBO signature.~ We note, however, that inefficient thermalization of the MHD magnetar wind within the ejecta can delay the rise of the SN lightcurve for up to one week, enhancing the visibility of the SBO peak   \cite{kasen2016magnetar}.~Inclusion of this effect in our calculations is currently under way.

A second feature is the impact of GW emission on lightcurves.~We find a moderate 
effect in the SBO emission, which becomes generally small in the ULTRASAT band.~As a consequence, the SBO lightcurves will not be sensitive to the potential GW emission by the central engine although, as already stated, they will allow the identification of a magnetar central engine over a sizeable range in parameter space.~On 
the other hand, we find that GW emission can induce a substantial modification in the UV (and bolometric) light-curve evolution of the SN.~This has two major implications:~first, it clarifies that
magnetar central engines do not necessarily produce super-luminous supernovae, since part of their spin energy could be radiated via GW.~A quantitative study of the  connection between these phenomena is postponed for future work.~Second, it shows 
the possibility to look for GW signatures in the SN light-curves alone, as their peak luminosities are reduced while their late-time decay can be significantly slower.~The latter, in particular, implies that magnetar-powered SN with GW emission from the central engine will be brighter, at $t>$ tens of days, than in the absence of GW emission.~A detailed investigation of this important result is also postponed for future work, and will entail a study of the whole parameter space, in particular disentangling the effects of changing the explosion energy ($E_{SN}$), the ejecta mass ($M_{\rm ej}$) and the opacity ($\kappa$).

The possibility to detect magnetar-powered SNe in larger numbers than their SBOs is particularly noteworthy for ULTRASAT.~Indeed, the SBO peak occurs typically within the first week of the explosion, and hence can only be well sampled if it occurs within the high-cadence region of the ULTRASAT survey.~On the contrary, the SN peaks and their late-time decays - possibly bearing the GW signatures - typically last several tens of days, and will thus be well-sampled even in the all-sky, low-cadence mode, further enhancing the detection rate. 

Within the estimated GW horizon ($> 4$ Mpc; see \cite{Sandhya23}) for the O5 science run, we expect ULTRASAT to easily detect all magnetar-driven SBOs, hence a multi-messenger detection rate $> 0.03$ yr$^{-1}$,~i.e.~a detection probability $> 8$\% assuming a 2.5 yr duration for O5.~The 7-8~times larger horizon foreseen for the Einstein Telescope will include the whole Virgo Cluster and ULTRASAT will be able to detect any SBOs within this distance.~Therefore, given a minimum estimated magnetar birth rate of $\sim$ 0.5 yr$^{-1}$ within 30~Mpc \cite{stella2005gravitational,dall2009early,Mereg24}, we expect a minimum rate of multi-messenger detections of 1 magnetar every 2 yrs. 

Our results demonstrate that magnetar-driven SBOs and their associated SNe will provide crucial EM triggers for GW searches of long-transients emitted by newly formed magnetars.~Besides signalling the start time for the transient GW, the light-curves of such events will help constrain the NS spin period and B-field, which in turn determine the GW signal shape.~Moreover, as highlighted above, ULTRASAT observations alone of magnetar-powered SNe may offer the unprecedented opportunity to pinpoint the emission of GWs by the central engine, informing direct GW searches in unique ways.


\section*{References}
\bibliography{moriond}

\end{document}